# Low temperature magnetic and dielectric properties of LnBaCuFeO$_5$ (Ln = Nd, Eu, Gd, Ho and Yb)

Surender Lal*, K. Mukherjee and C. S. Yadav

School of Basic Sciences, Indian Institute of Technology Mandi, Mandi 175005, Himachal Pradesh, India

*Corresponding Author: Surender Lal

School of Basic Sciences,

Indian Institute of Technology Mandi,

Mandi 175005, Himachal Pradesh, India

Tel: +919418304396

E-mail: surenderlal30@gmail.com

## Abstract

The layered perovskite compounds are interesting due to their intriguing physical properties. In this article we report the structural, magnetic and dielectric properties of $LnBaCuFeO_5$ (Ln=Nd, Eu, Gd, Ho and Yb). The structural parameters decrease from Nd to Yb due to the decrease in the ionic radii of the rare earth ions. An antiferromagnetic transition is observed for $EuBaCuFeO_5$ near 120 K along with the glassy dynamics of the electric dipoles below 100 K. The magnetic transition is absent in other compounds, which may be due to the dominance of the magnetic moment of the rare earth ions. The dielectric constant does not show any anomaly, except in the case of $HoBaCuFeO_5$ where it shows a weak frequency dependence around 54 K. These compounds show a significant enhancement of dielectric constant at high temperatures which have been attributed to Maxwell-Wagner effect. However, no significant magneto-dielectric coupling has been observed in these layered perovskites.

Keywords: Layered perovskites, Dielectric constant, Maxwell Wagner Polarization, Magneto-dielectric coupling.

## 1. Introduction

Layered perovskite materials of the form AA'BB'O$_5$, where A, A' are alkali or rare-earth metals and B, B' are transition metals with same or different ions are under extensive investigation in recent years [1–6]. These materials are important as they exhibit multiferroicity near room temperature and also due to the fact that the ferroelectricity in these compounds is induced by magnetic ordering [4–9]. The compound YBaCuFeO$_5$ belongs to this family and it undergoes both magnetic and dielectric transitions above 200 K [4, 6, 9]. In this compound a paramagnetic to commensurate antiferromagnetic (C-AFM) ordering is noted at ~ 440 K followed by a changeover from C-AFM to incommensurate antiferromagnetic (IC-AFM) state at ~ 200 K [4,6,9]. The magnetic and the dielectric transitions of these layered perovskites can be tuned by changing the chemical pressure or by introducing a disorder [1,4,6,8–12]. Various rare-earth ions at the A-site can change the chemical pressure and it may result in interesting dielectric properties in the respective compounds. A detailed comparative study of the contrasting physical properties exhibited by rare earth layered perovskite LaBaCuFeO$_5$ and LuBaCuFeO$_5$ has been carried out [13]. Recent investigations on HoBaCuFeO$_5$, GdBaCuFeO$_5$ and YbBaCuFeO$_5$ reveal that the observed upturn in the heat capacity at low temperatures in these compounds is attributed to the Schottky anomaly [14,15]. However, to the best of our knowledge there are few reports on low temperature dielectric properties of other member of this family.

In this work the structural and dielectric properties of rare-earth based layered perovskite compounds LnBaCuFeO$_5$ (where Ln = Eu, Nd, Yb, Gd and Ho) have been investigated. Along with it magnetic studies of EuBaCuFeO$_5$ and NdBaCuFeO$_5$ are also reported. The key observations of this work are: i) EuBaCuFeO$_5$ undergo C-AFM to IC-AFM ordering around 120 K. Glassy dynamics of the electric dipoles is noted at low temperature region. ii) No magnetic transition is noted for NdBaCuFeO$_5$. iii) Dielectric behavior of NdBaCuFeO$_5$, YbBaCuFeO$_5$ and GdBaCuFeO$_5$ in the low temperature region is featureless. However, in HoBaCuFeO$_5$ a frequency dependent anomaly is noted in the low temperature region. But the interactions between the electric dipoles are significantly weaker which results in the absence of glassy behavior. iv) In all these compounds the dielectric constant significantly increases at high temperatures and this feature have been is attributed to the Maxwell Wagner effect. vi) Magneto-dielectric coupling is very weak in all these compounds.

## 2. Experimental details

The compounds GdBaCuFeO$_5$, HoBaCuFeO$_5$ and YbBaCuFeO$_5$ are the same as studied in Ref [14]. The compound EuBaCuFeO$_5$ is the same as studied in Ref [15]. The compound NdBaCuFeO$_5$ is prepared under similar condition as mention in Ref [9]. Power x-ray diffraction was performed using Rigaku smart lab diffractometer using monochromatized Cu K$\alpha_1$ radiation at room temperature. The DC magnetization measurements were carried out with a SQUID magnetometer (Quantum Design USA) in the temperature range of 2-350 K. Hioki LCR meter was used to for temperature and magnetic field dependence of dielectric constant measurements setup from Cryonano lab, integrated with the PPMS.

## 3. Results and discussion

### 3.1 Structural Properties of LnBaCuFeO$_5$ (Ln = Nd, Eu, Gd, Ho and Yb)

The structural analysis of room temperature x-ray diffraction data have been carried out for the series of compounds LnBaCuFeO$_5$ (Ln=Nd, Eu, Gd, Ho and Yb) through Rietveld refinements. The obtained parameters are tabulated in Table 1. For GdBaCuFeO$_5$, HoBaCuFeO$_5$, YbBaCuFeO$_5$ and EuBaCuFeO$_5$; some parameters are taken from Ref [14] and [15], while for YBaCuFeO$_5$, LaBaCuFeO$_5$ and LuBaCuFeO$_5$ some parameters are taken from Ref [9] and [13] for the sake of comparison. LnBaCuFeO$_5$ are layered perovskite compound with tetragonal structure of P4mm space group [1,2,16,17]. The structure of LnBaCuFeO$_5$ can be described as consisting of [CuFeO$_5$]$_\infty$ double layers of the corner sharing CuO$_5$ and FeO$_5$ pyramids along c-direction, containing Ba$^{2+}$ ions [1,10,11]. These [CuFeO$_5$]$_\infty$ double layers are separated by Ln$^{3+}$ planes. The CuO$_5$/FeO$_5$ forms the pyramids which are linked together via apical oxygen to form bipyramid and are aligned ferromagnetically within the bipyramids. However, these bipyramids are aligned anti-ferromagnetically along the c-direction, which gives rise to spiral antiferromagnetic magnetic ordering in this direction at low temperature in case of YBaCuFeO$_5$ [6]. The change in the structural properties has the influence on the magnetic and dielectric properties of the compounds [7,9].

From the structural analysis it is observed that inter-pyramid distance decreases with lanthanide contraction whereas the distance of Cu/Fe from the apical oxygen goes on increasing from La to Lu. It is also noted that the rare-earth ions with higher ionic radii leads to the expansion in the unit cell whereas those with lower ionic radii leads to compression in the unit cell as compared with the unit cell volume of YBaCuFeO$_5$ [14]. The lattice parameters of the unit cell decreases from Nd to Yb due to the decrease in the ionic radii of the rare earth ions following the lanthanide contraction rule. This leads to change in the lattice parameters such as bond length, bond angle and the inter-pyramid distance, which in turn influences the magnetic and the dielectric properties of the compounds. In the following sections we present the results observed for different members of LnBaCuFeO$_5$.

3.2 EuBaFeCuO$_5$

Figure 1 shows the temperature dependent zero field cooled (ZFC) and field cooled (FC) DC magnetic susceptibility of EuBaCuFeO$_5$ in the temperature range of 2-350 K measured at 5 kOe. It is noted that the curve shows a broad AFM transition near 120 K. The magnetization increases almost linearly with the decrease in temperature down to 120 K. This transition is analogous to the C-AFM to IC-AFM transition observed at 200 K in YBaCuFeO$_5$. The bifurcation between the ZFC and FC is absent in the entire temperature range of measurement. The antiferromagnetic nature of the compounds can be further confirmed from the field dependence of isothermal magnetization data as reported in Ref [15], where the linear behavior along with the absence of hysteresis is noted. We would like to mention that in this compound the spiral nature of the magnetic ion is along the *b*-direction and the commensurate-incommensurate nature is only observed along the *c*-direction [6]. The interaction along the *c*-direction depends upon the distance between the magnetic ions. Here the presence of Eu in place of Y increases the unit cell volume and the lattice parameter *c*. This increase along the *c*-direction moves the bipyramids apart reducing the antiferromagnetic interaction among the magnetic ions. This results in shifting of the C-AFM to IC-AFM transition towards lower temperature. Figure 2 a and b shows the temperature dependence of real ($\varepsilon'$) and imaginary ($\varepsilon''$) part of dielectric constant measured in the temperature range of 10 to 300 K for EuBaCuFeO$_5$ at different frequencies. As noted from figure 2 (a), $\varepsilon'$ is independent of frequency upto 120 K, beyond which the curves show frequency dependence. Corresponding $\varepsilon''$ also shows similar features. However, around this low temperature region (40 -90 K), a broad frequency dependent peak is noted. To crosscheck this anomaly $\varepsilon'$, the derivative of the $\varepsilon'$ is plotted as a function of temperature upto 100 K and shown in the inset of the figure 2 (a). At 10 kHz, d$\varepsilon'$/d$T$ shows a peak around 54 K. The peak temperature increases as the frequency is increased. The observation of such features indicates the presence of glassy dynamics of the electric dipoles. To shed some more light about the dynamics, the variation of the peak temperature is analyzed by the Vogel Fulcher (VF) law [18]:

$$\tau = \tau_0 \exp\left[\frac{E_a}{K_B(T_f - T_0)}\right] \quad \ldots (1)$$

Where $T_0$ is the measure of the interaction strength between the electric dipoles and $E_a$ is the average activation energy. Lower inset of figure 2(b) shows the fitting of VF law in the range 10 to 300 kHz. The parameters $\tau_0$ and $T_0$ obtained from the fitting are $3.5 \times 10^{-8}$ sec and 36 K respectively and $E_a$ is ~0.01 eV. From the parameters, the value of $E_a/k_B$ is obtained to be 120 K. Hence, in this case $T_0 < E_a/k_B$, which is a signature of weak coupling among the electric dipoles [19]. Hence, it can be said that in EuBaCuFeO$_5$ there is a presence of glassy electric-dipole dynamics below 100 K. Above 200 K, the magnitude of both ε' and ε'' increases significantly. The appearance of dielectric transition close to the magnetic ordering temperature indicates the possibility of the coupling between the magnetic and electric order parameters. To check the presence of the magneto-dielectric (MDE) coupling we have measured ε' as a function of magnetic field at two temperatures: one in the glassy phase and one above it. Figure 2 (c) shows MDE (calculated using the formula $\Delta\varepsilon'(\%) = [(\varepsilon'_H - \varepsilon'_{H=0})/\varepsilon'_{H=0}] \times 100$, where $\varepsilon'_H$ and $\varepsilon'_{H=0}$ are the dielectric constant with and without magnetic field respectively) as a function of magnetic field at 10 kHz at 70, 100 and 200 K. At 100 K a weak coupling is noted while at other temperatures MDE coupling is absent.

3.3 NdBaCuFeO$_5$ and YbBaCuFeO$_5$

The temperature dependent DC magnetic susceptibility of NdBaCuFeO$_5$ under ZFC and FC condition is carried out at 5 kOe (shown in figure 3). No magnetic transition is observed down to 2 K and both ZFC and FC curves superimposes on each other. The temperature response of inverse susceptibility curve is fitted with Curie Weiss law in the temperature range of 2 -390 K. The obtained effective moment ($\mu_{eff}$) and the Curie Weiss temperature ($\theta_{cw}$) are ~ 2.34 $\mu_B$ and -339 K respectively. The negative value of $\theta_{cw}$ indicates the dominance of antiferromagnetic exchange interactions in this compound. The absence of any magnetic transition in this compound is possibly due to the fact that the moments of the Nd ion dominate the effect arising out of Cu/Fe ions. The temperature response of magnetization of YbBaCuFeO$_5$ also shows similar behavior and has already been reported in Ref [14].

Figure 4 (a) and (b) shows $\varepsilon'$ as function of temperature for NdBaCuFeO$_5$ and YbBaCuFeO$_5$. For both the compounds it is noted that $\varepsilon'$ is independent of frequency upto 100 K. The corresponding $\varepsilon''$ of the compounds is also featureless in the low temperature region, as is seen from figure 4 (c) and (d). Also no anomaly is observed in this temperature range which is confirmed from temperature response of $\varepsilon''$ (Figure 4 (b) and (e)). A sharp increase in $\varepsilon'$ and $\varepsilon''$ is observed above 100 K for both these compounds. The temperature dependent $\varepsilon''$ for NdBaCuFeO$_5$ and YbBaCuFeO$_5$ shows a peak around 188 K and 138 K respectively at 5 kHz. For both these compounds the peak shifts towards higher temperature on increasing the frequency. Here we would like to mention that the loss factor above 100 K is quite significant for both these compounds. To check the possibility of the presence of MDE coupling in these compounds, $\Delta\varepsilon'$ is measured as a function of magnetic field at selected temperatures (shown in Figure 4 (e) and (f)). For NdBaCuFeO$_5$, $\Delta\varepsilon'$ data is not very clean but shows small ($< 0.5\%$), field independent value for the measured temperatures. For YbBaCuFeO$_5$, $\Delta\varepsilon'$ shows field dependence (although with small value $\Delta\varepsilon' < 1\%$). It is interesting to note that $\Delta\varepsilon'$ is negative for 50 and 200 K whereas it is positive for 100 K. This behavior is similar to that observed for YBaCuFeO$_5$ [9].

3.4 GdBaCuFeO$_5$ and HoBaCuFeO$_5$

The temperature response of the magnetization of GdBaCuFeO$_5$ and HoBaCuFeO$_5$ do not show any transition down to 1.8 K and has been already reported in Ref [14]. Figure 5 (a) and (c) shows the temperature response of $\varepsilon'$ and $\varepsilon''$ for GdBaCuFeO$_5$ in the range 10 to 300 K at different frequencies. It is noted that the temperature dependence of $\varepsilon'$ is dispersion less 200 K. Above 200 K, $\varepsilon'$ (and $\varepsilon''$) increases sharply and shows the dispersion with increase in the frequency. Figure 5 (b) and (d) shows the variation of $\varepsilon'$ and $\varepsilon''$ with temperature at different frequencies for HoBaCuFeO$_5$. Below 175 K, $\varepsilon'$ is featureless, but shows weak frequency dependence. However, a broad peak is around 50 K at 10 kHz is noted in the temperature response of $\varepsilon''$ (inset of figure 5 (d)). With increase in frequency, peak temperature shifts to the higher temperature side. The presence of frequency dependence indicates the presence of some glassy dynamics of the electric dipoles in the low temperature region of this compound, similar to that observed for EuBaCuFeO$_5$. This observation is analyzed with Arrhenius and VF law. It is noted that for HoBaCuFeO$_5$, better fit is obtained with Arrhenius law, in contrast to that observed for EuBaCuFeO$_5$. Inset of figure 5 (d) shows the fitting of the variation of the peak temperature with Arrhenius law of the form [21]:

$$\tau = \tau_0 \exp\left[\frac{E_a}{K_B T}\right] \quad \ldots (2)$$

where $\tau_0$ is the pre-exponential factor, $E_a$ denotes the activation energy, $k_B$ is the Boltzmann constant. From the fitting we obtained $E_a = 0.037$ eV and the $\tau_0 = 9.6 \times 10^{-10}$ sec. This indicates that the observed anomaly in HoBaCuFeO$_5$ is due to the thermally activated dipoles. However, the interaction between the dipoles is not strong enough resulting in the absence of collective freezing. Above 200 K, a sharp increase in $\varepsilon'$ and $\varepsilon''$ has been observed for both these compounds. The temperature dependent $\varepsilon''$ shows a peak around 200 K for 5 kHz, which shifts towards higher temperature on increasing the frequency. Figure 5 (e and f) shows the magnetic field dependence of $\Delta\varepsilon'(\%)$ measured at different temperatures. The magnetic field dependence of dielectric constant shows the non-zero values in the temperature range of 100 - 200 K. The presence of the very small value of $\Delta\varepsilon'$ indicates poor magnetodielectric coupling in HoBaCuFeO$_5$ and GdBaCuFeO$_5$.

## 4. Discussion

The magnetic studies of LnBaCuFeO$_5$ (Ln = Nd, Eu, Gd, Ho, Yb); show the absence of magnetic ordering, except for EuBaCuFeO$_5$. It is possible that the effective moment of the rare earth ions influences the FM and AFM coupling of the transition metal ions and suppresses this ordering [14]. Since Eu$^{3+}$ ion is non-magnetic, the magnetic ordering (C-AFM to IC-AFM) similar to YBaCuFeO$_5$ remains intact [4,6,9]. EuBaCuFeO$_5$ has larger unit cell volume (larger c-parameter) in comparison to YBaCuFeO$_5$, which results in reduced size of Cu/FeO$_5$ bipyramid (enhanced FM coupling) and enhanced inter-bipyramidal distance (reduced AFM coupling). This variation in FM-AFM magnetic interaction leads to the reduction in C-AFM to IC-AFM transition temperature.

The low temperature dielectric behavior is featureless in all the compounds except for EuBaCuFeO$_5$ and HoBaCuFeO$_5$. These two compounds show frequency dependent anomaly at low temperatures. Analysis of the shift in the peak temperature of the dielectric curve for EuBaCuFeO$_5$ indicates that freezing of dipole results in dipolar glass like state as the temperature is lowered. But for HoBaCuFeO$_5$, relatively weaker interaction between the dipoles results in the absence of collective freezing mechanism. At high temperatures, the dielectric constant of all these layered perovskites increases significantly above 200 K. However, it is to note that the loss factor is quite high in this temperature range. It is possible that the dielectric relaxation observed in these polycrystalline compounds is due to the presence of inhomogeneity resulting in different conductivity within the grains and at the grain boundaries. The charge inside the grains is trapped by the high potential of the grain boundaries and it acts as the thin capacitors resulting in the net capacitance. This non-intrinsic contribution to dielectric constant along with the high value of the dielectric loss can be attributed to the Maxwell Wagner (MW) effect [20].

Low value of Δε' for these compounds indicates poor MDE coupling. The magnetoelectric coupling is not the only way to produce MDE effect; other mechanism such as the magneto-resistance, MW polarization can also give rise to MDE effect. There are certain materials in which the MDE effect is found but the spontaneous polarization is not observed [8,12,21]. Some materials show the MDE coupling due to the combined effect of magneto-resistance and MW effect [21]. In the absence of the magnetic transition the MDE effect maybe due to the extrinsic effect such as MW polarization that induces this effect [22]. Similar behavior in these layered perovskites suggests the contribution of MW polarization in the dielectric behavior. The weak MDE coupling in some of these compounds is due to the effect of MW polarization and not due to the magnetoelectric coupling [21].

## 5. Conclusions

We have investigated the physical properties of rare earth layered perovskite compounds LnBaCuFeO$_5$ (Ln = Eu, Nd, Yb, Gd and Ho). A magnetic transition is noted only for EuBaCuFeO$_5$ while it is absent in the other studied compounds. The glassy dynamics of electric dipoles is observed in the low temperature region of EuBaCuFeO$_5$ compound. However, in HoBaCuFeO$_5$ interactions between the dipoles is not strong enough resulting in the absence of collective freezing of dipoles in the low temperature region. The dielectric behavior of the other compounds is featureless at low temperatures. These compounds show a significant enhancement of dielectric constant at high temperatures which have been attributed to Maxwell-Wagner effect. No significant magneto-dielectric coupling has been observed in these compounds.

**Acknowledgement**

The authors acknowledge IIT Mandi for providing the experimental facilities. SL acknowledges the UGC India for SRF Fellowship.

Table 1: Lattice parameters calculated from the Rietveld refinement of x-diffraction of LnBaCuFeO$_5$ (Ln = Nd, Eu, Gd, Ho, Y, Yb)

| Sample | La Ref[13] | Nd | Eu | Gd | Ho | Y Ref [9] | Yb | Lu Ref[13] |
|---|---|---|---|---|---|---|---|---|
| **Radii (Å)** | 1.032 | 0.983 | 0.947 | 0.938 | 0.901 | 0.9 | 0.86 | 0.861 |
| **a (Å)** | 3.933(2) | 3.923 | 3.899 | 3.893(2) | 3.875(1) | 3.871(0) | 3.855(2) | 3.857 (2) |
| **c (Å)** | 7.825(0) | 7.755(0) | 7.703(1) | 7.685(1) | 7.661(3) | 7.663 (1) | 7.636(1) | 7.646 (1) |
| **volume(Å$^3$)** | 121.01 | 119.32 | 117.10 | 116.47 | 115.035 | 114.83 | 113.48 | 113.79 |
| $\chi^2$ | 1.59 | 1.70 | 1.86 | 2.32 | 1.90 | 2.18 | 5.48 | 1.53 |
| **R-factor** | 9.35 | 8.59 | 4.84 | 6.06 | 4.47 | 4.92 | 13.88 | 3.72 |
| **RF- Factor** | 7.45 | 9.60 | 4.40 | 6.44 | 4.61 | 4.18 | 12.59 | 4.26 |
| **Inter-pyramid distance(Å)** | 3.613(1) | 3.580(0) | 2.848(2) | 2.841(1) | 2.832(2) | 2.833(3) | 2.824(1) | 2.827 (2) |
| **Fe-O1 (Å)** | 1.989 | 1.971 | 2.128 | 2.126 | 2.050 | 2.117 | 2.110 | 2.112 |
| **Cu-O2 (Å)** | 1.861 | 1.845 | 1.994 | 1.989 | 2.050 | 2.048 | 1.977 | 2.043 |
| **Fe-O3 (Å)** | 1.974 | 1.969 | 1.969 | 1.966 | 1.957 | 1.955 | 1.948 | 1.948 |
| **Cu-O4 (Å)** | 2.037 | 2.031 | 2.002 | 1.999 | 1.990 | 1.975 | 1.981 | 1.968 |
| **Fe-Cu distance(Å)** | 3.975 | 3.939 | 3.581 | 3.572 | 3.561 | 3.499 | 3.551 | 3.491 |

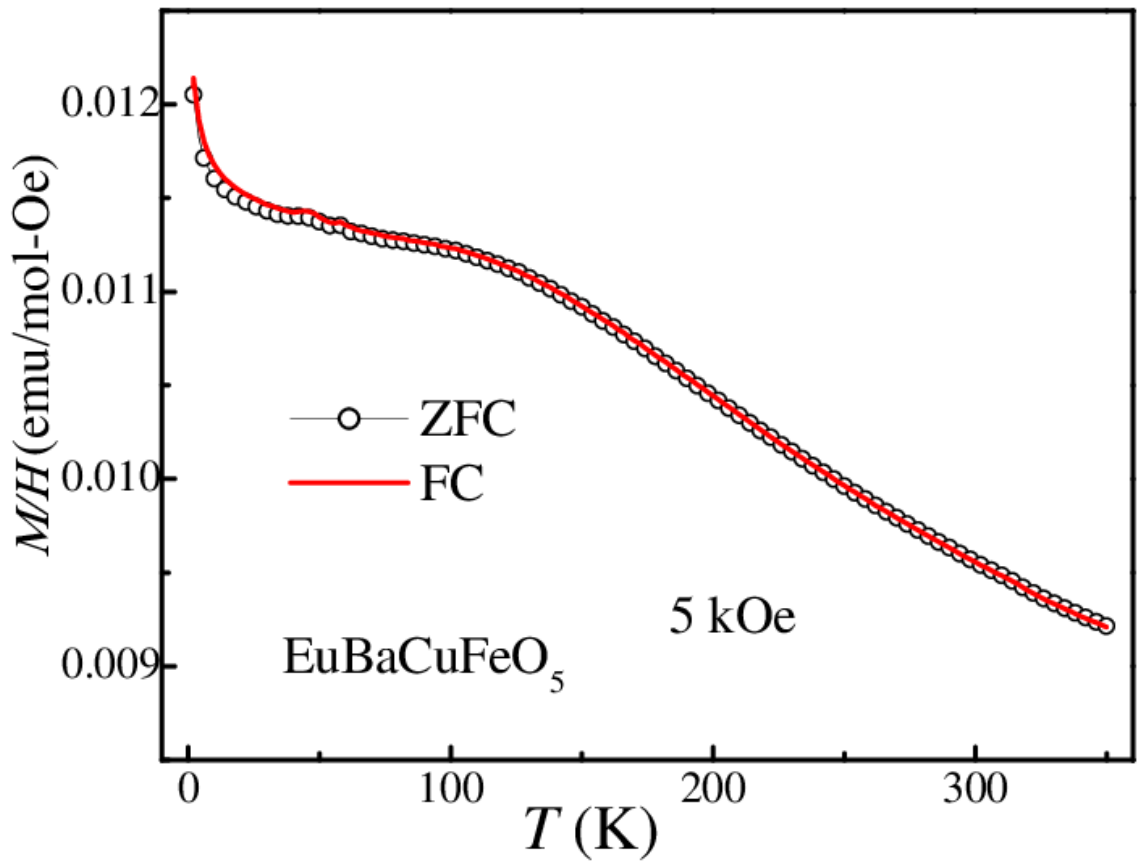

Figure 1: Temperature dependence of DC susceptibility of EuBaCuFeO$_5$ measured at 5 kOe magnetic field.

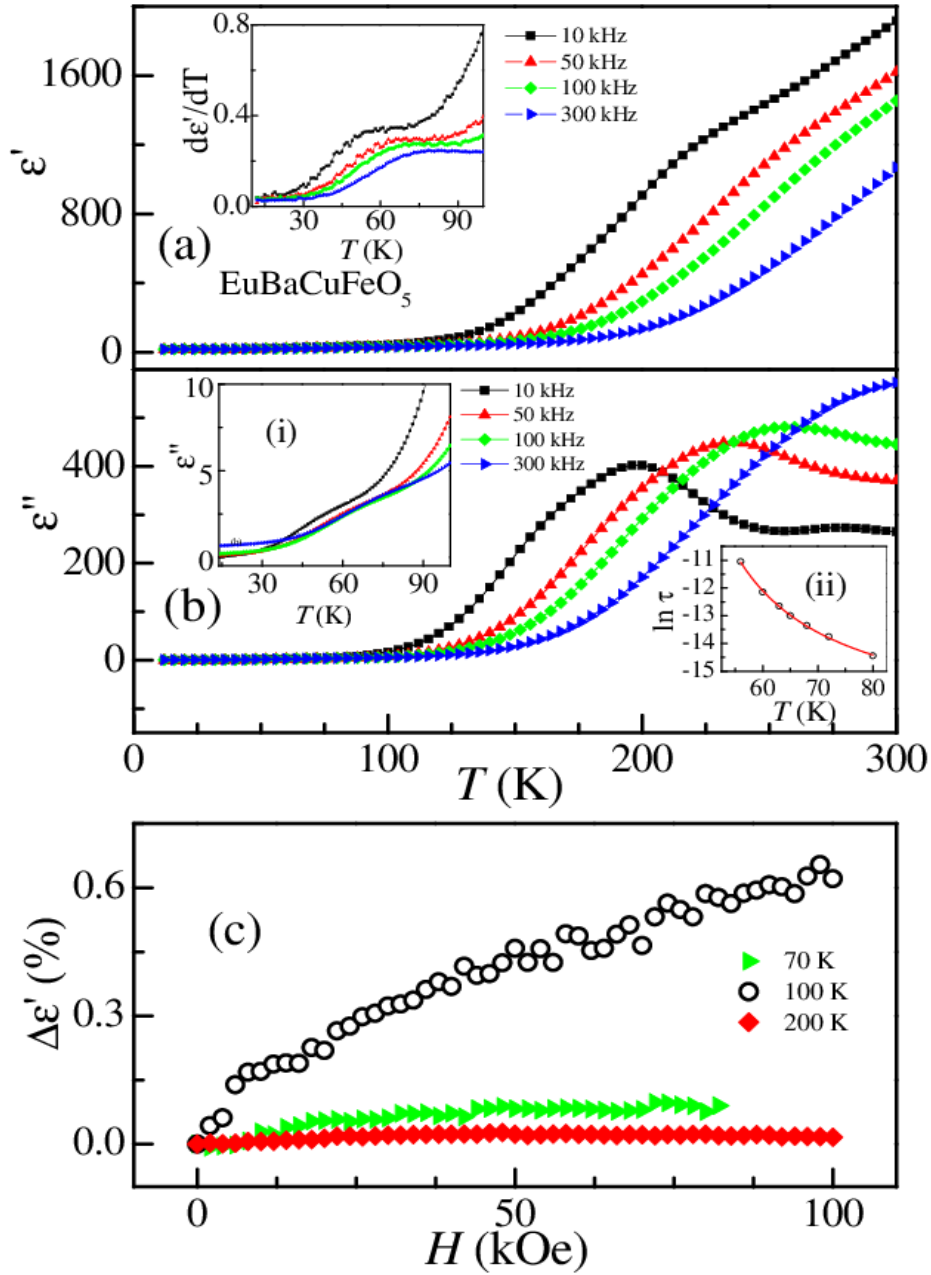

Figure 2 (a) Temperature dependence of real (a) and imaginary (b) parts of dielectric constant of EuBaCuFeO$_5$ measured at $f$ =10, 50, 100 and 300 kHz. Inset of (a) shows dε'/dT plots as a function of temperature at low temperature. Insets of (b) shows (i) Imaginary part of the dielectric constant at low temperature 20 to 100 K and (ii) Vogel Fulcher fit of the peak temperature respectively. (c) Relative change in the dielectric constant under the application of magnetic field measured at 10 kHz.

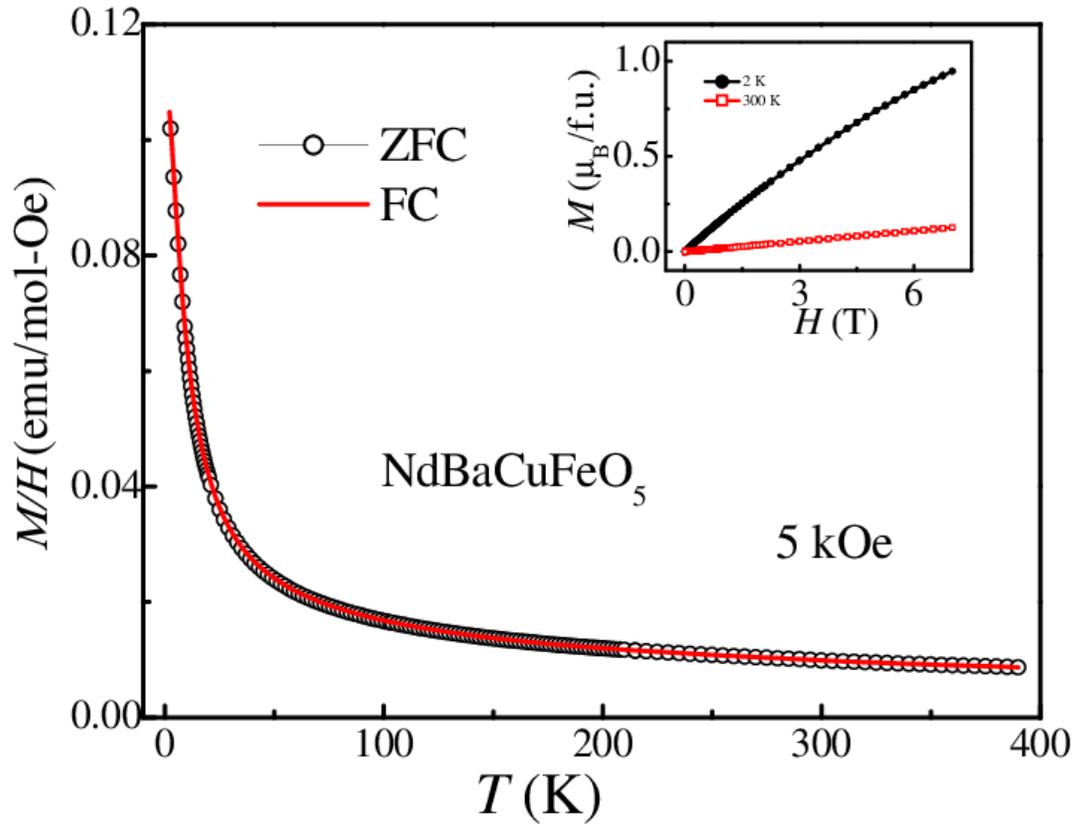

Figure 3: Temperature dependence of DC susceptibility of NdBaCuFeO$_5$ measured at 5 kOe magnetic field. Inset shows the isothermal magnetization measured at 2 and 300 K up to 7 T magnetic field.

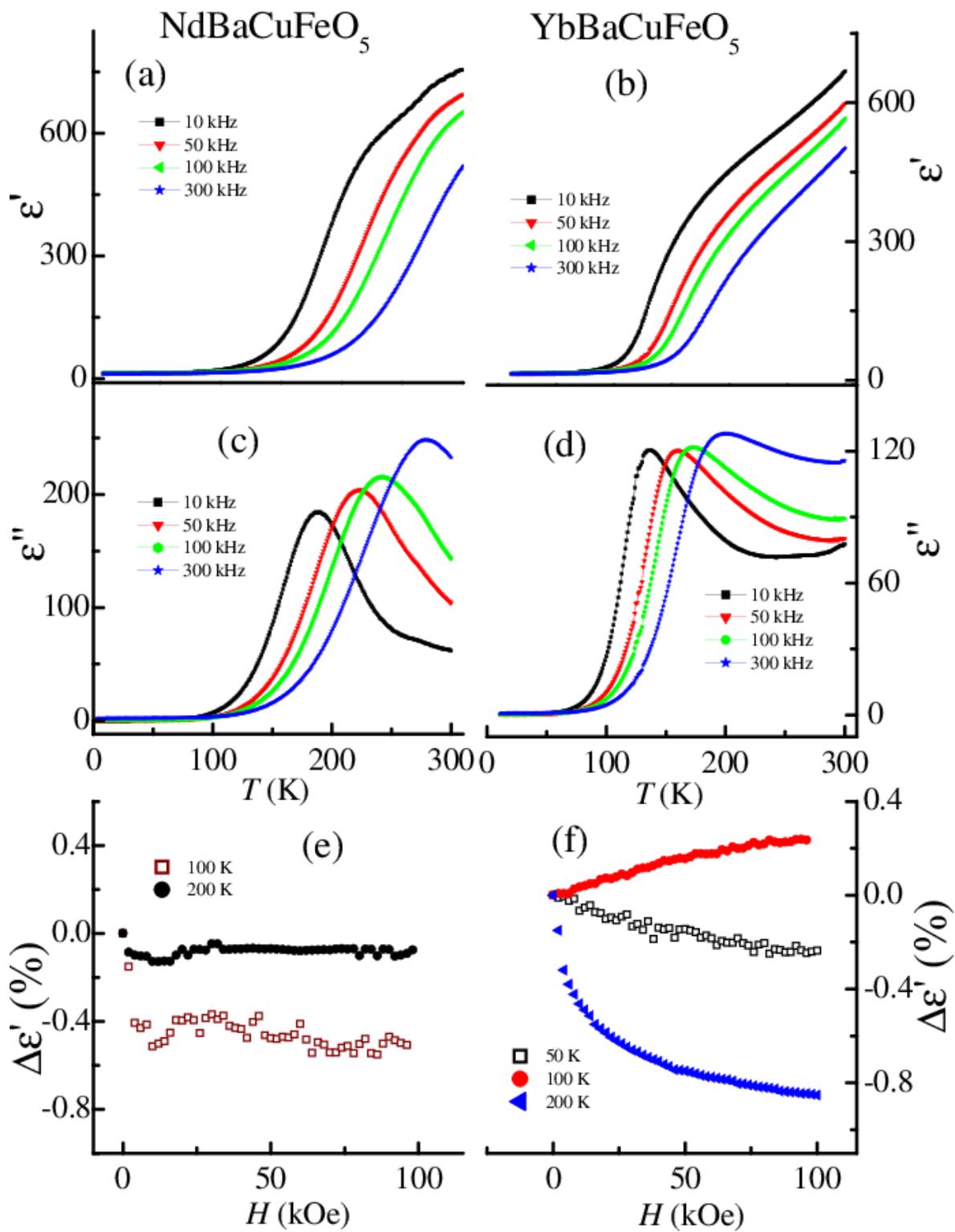

Figure 4: Temperature dependence of real(a, b) and imaginary (c, d) parts of dielectric constant of $NdBaCuFeO_5$ and $YbBaCuFeO_5$ measured at different frequency. (e) and (f) Magnetic field dependence of relative dielectric permittivity of $NdBaCuFeO_5$ and $YbBaCuFeO_5$ measured at selected temperatures at frequency of 10 kHz.

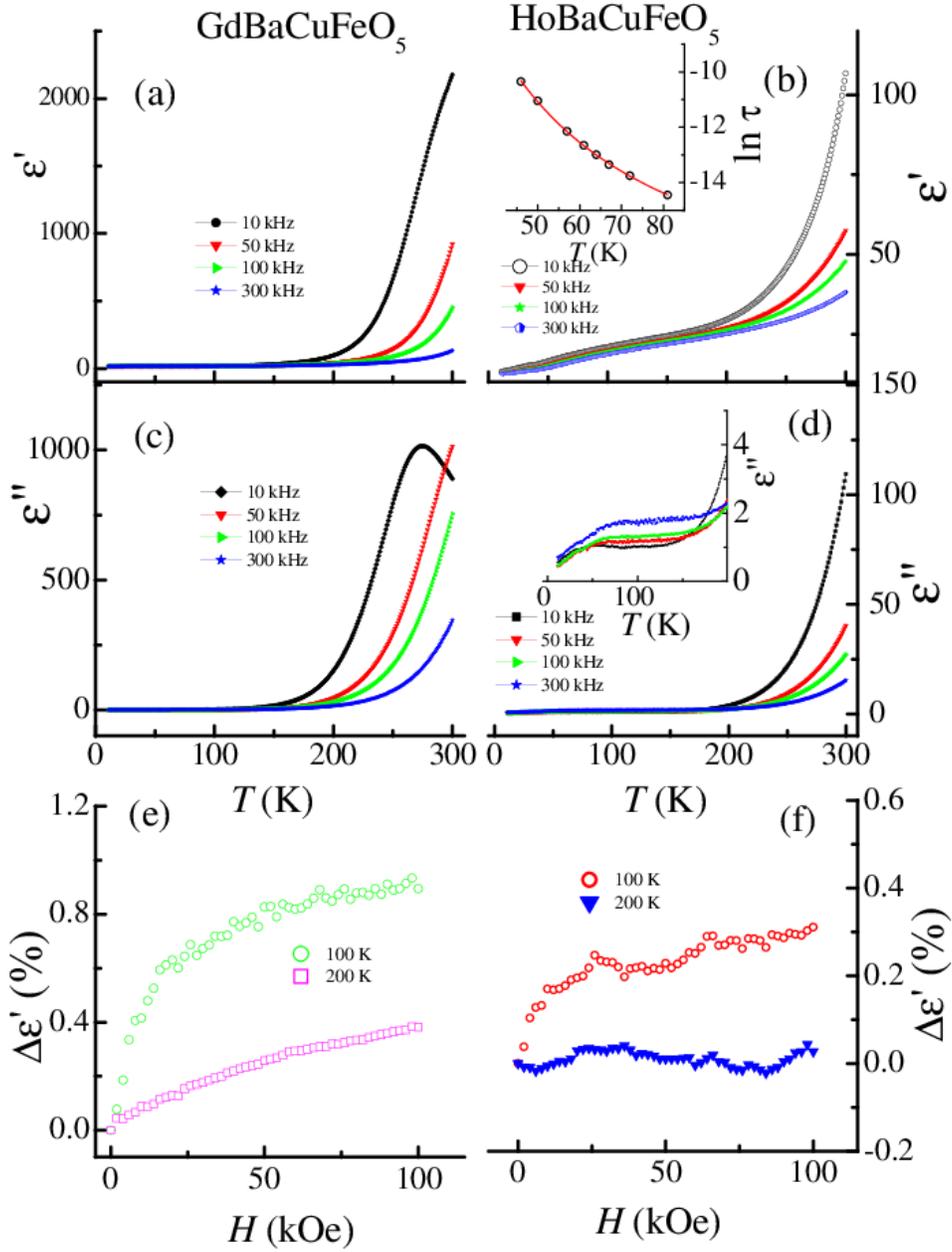

Figure 5: Temperature dependence of the real (a, b) and imaginary (c, d) parts of the dielectric constant of GdBaCuFeO$_5$ and HoBaCuFeO$_5$ measured at different frequency. Inset of (b) shows the Arrhenius fit of the peak temperature. Inset of (d) shows the imaginary part of dielectric constant at low temperature. (e) and (f) Magnetic field dependence of relative dielectric permittivity of GdBaCuFeO$_5$ and HoBaCuFeO$_5$ measured at 100, 200 K at $f =$ 10 kHz.